# Network Theory, Cracking and Frictional Sliding

Ghaffari, H.O.
*Department of Civil Engineering and Lassonde Institute, University of Toronto, Toronto, Canada*
Young, R.P.
*Department of Civil Engineering and Lassonde Institute, University of Toronto, Toronto, Canada*

**ABSTRACT:** We have developed different network approaches to complex patterns of frictional interfaces (contact areas developments). Network theory is a fundamental tool for the modern understanding of complex systems in which, by a simple graph representation, the elementary units of a system become nodes and their mutual interactions become links. With this transformation of a system to network space, many properties about the structure and dynamics of the system itself can be inferred. We map the real-time net contact areas to network configurations while we use similarity measures to link the nodes. In other words, we follow the possible collective deformation of contact areas as well as the characteristics of correlated elements. Here, we analyze the dynamics of static friction. We found, under the correlation measure, the fraction of triangles correlates with the detachment fronts. Also, for all types of the loops (such as triangles), there is a universal power law between nodes' degree and motifs where motifs frequency follow a power law. This shows high energy localization is characterized by fast variation of the loops fraction. Also, this proves that the congestion of loops occurs around hubs. Furthermore, the motif distributions and modularity space of networks –in terms of within-module degree and participation coefficient- show universal trends, indicating an in common aspect of energy flow in shear ruptures. Moreover, we confirmed that slow ruptures generally hold small localization, while regular ruptures carry a high level of energy localization. We proposed that assortativity, as an index to correlation of node's degree, can uncover acoustic features of the interfaces. We showed that increasing assortativity induces a nearly silent period of fault's activities. Also, we proposed that slow ruptures resulted from within-module developments rather than extra-modules of the networks. Our approach presents a completely new perspective of the evolution of shear ruptures.

## 1. INTRODUCTION

The evolution of frictional interfaces and the onset of slip are integrated with the developments of the contact areas [1-5, 10, 12-15, 29]. The crack-like behavior of rupture in frictional interfaces also supports the role of relative contact areas and apertures. Crack-like behavior of shear rupture (or generally fracture type 2) is accompanied by transient behavior of the fracture. The transition from static to dynamic friction with regard to rupture nucleation and precursors is the key feature in the sliding process of frictional interfaces. A basic process in transition from slip to sliding state (stick-slip) includes the propagation of detachment fronts where reduction of the contact areas yields fast energy emission. The initiation of detachment fronts is followed by the emission of acoustic waves, the main tool in our understanding from earthquakes. Detachment fronts (front-like ruptures) are the wave-like fronts that are formed during the local and global fast change of contact areas, crossed or arrested through the interface [1-3]. By employing recent advancements in data acquisitions systems, laboratory experiments reveal three rupture modes [1-7]: slow ruptures, sub-Rayleigh ruptures (well known as regular earthquakes) and super-shear rupture. The formation, transition and arresting of shear rupture modes—as well as detachment fronts which are mostly sub-Rayleigh forms— are the most elusive problems in terms of loading configurations and the geometrical complexity of frictional interfaces. Although different numerical and analytical methods investigated several aspects of regular and super-shear ruptures [8-10], the inherent complexities embedded in fault surfaces and possible collective deformation of an interface's elements necessitates further detailed analysis of shear ruptures. Recent experimental observations [1-5] show a clear pattern of slow ruptures, as well as transition of regular ruptures to slow fronts; slow ruptures to sub-Rayleigh fronts; or the arresting of the rupture fronts. From another perspective, the careful investigation of particles' correlation patterns in sheared materials –using numerical and experimental evidence [16-17] – showed the emergence of anisotropic long-range correlated patterns during the deformation of the sheared system. It has been shown that long-range correlations in the fluctuations of shear strain induce the excitation of additional elastic modes. The strong correlation in the direction of the shear lowers the effective resistance to rupture in the direction of the shear [17].

In this study, we investigate the possible dependency of shear rupture transitions with correlation patterns. To characterize the similarity of real-time friction patterns, we map the interface's elements into the proper networks. With this transformation, the networks' parameters are related to the state variable in friction laws, such that the evolution of frictional interfaces can be modeled in terms of the networks' evolution. Remarkably, analysis of different real-time contact measures showed that the rupture transitions strongly scaled with the motif evolution of networks as well as the fraction of triangles (clustering coefficient). Furthermore, we found that emerged assortative networks show a unique power law scaling in terms of global and local similarities. Interestingly, irrespective of rupture speed, similar trends in the motifs' ranks are observed. Our findings about the modularity of friction-networks with respect to the within-module degree (Z) and participation coefficient (P) indicate that evolvable frictional interfaces occupy certain regions of the modularity space. We show that suddenly increasing assortativity induces a nearly silent period of the fault's activities. To investigate the possible coupling between the dynamic and topology of contact areas, we identified contact-patches that have many interacting partners ("hubs") in a network of contact patches' interactions (here after friction network) selected from recent experiments. Based on our solid results, we proposed that slow ruptures resulted from hubs' activity within modules rather than the hubs' interaction in the networks' extra-modules. This approach can facilitate more realistic modeling of acoustic waveforms.

## 2. DATA AND METHODOLOGY

Our data set includes the developments of real-time contact areas in recent friction experiments on transparent interfaces [1, 3-4]. Generally, the experiments involve shearing two transparent blocks on each other while a uniform normal loading is on the top block and a tangential force is used in the trailing edge (for more details of the experiments see the references). Recordings of the real-time relative contact areas mostly are based on a 1D assumption of interface dimension. It considers average optical intensity by employing a laser ray through the interface [1, 4]. However, there are a few cases of 2D measures of relative contact areas corresponding to the intensity of laser-light. We use correlation and Euclidean metrics over 2D relative contact patterns (figure 1a). Furthermore, for the 1D interface, we employ a truncated norm and a standard-delay-coordinate embedding method to analyze the time-series and compare the similarity of the patterns. To set up a non-directed network over 2D contact areas in a certain time step, we considered each patch of measured contact areas perpendicular to shear direction as a node – Figure.2a.

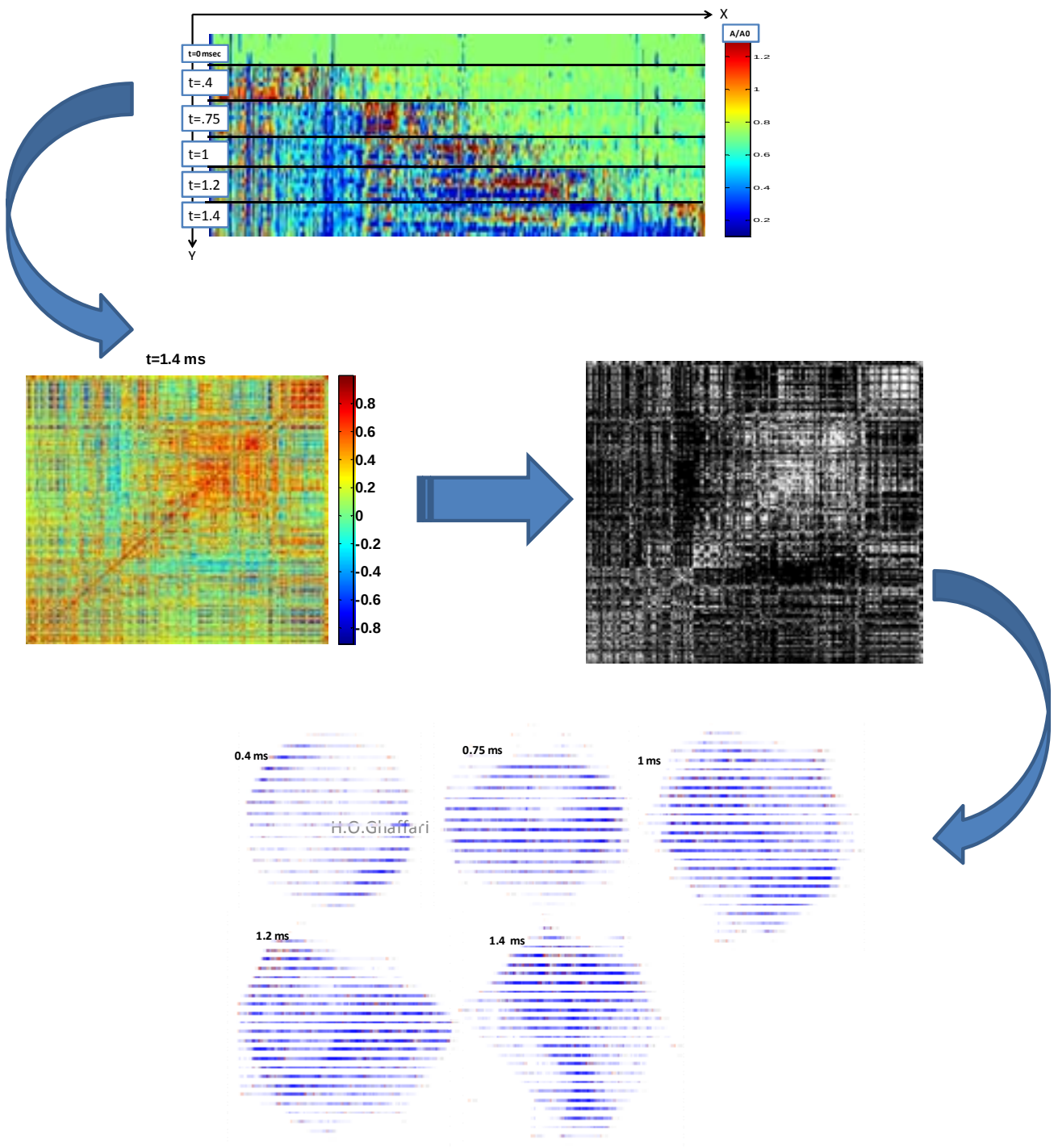

Fig.1. Methodology to extract friction networks from real t-time contacts: Evolution of real-time contacts through 6 time-windows in X-Y space of an interface. Each pixel corresponds with net contact area (contact areas data set courtesy of J. Fineberg); visualization of the correlation matrix and achieved friction networks through 5 time-windows.

Each profile has $N$ pixels where each pixel shows the relative-contact area of that cell. Then, we define correlations in the profiles by using:

$$C_{ij} = \frac{\sum_{l=1}^{N}\left[A_i(l) - \langle A_i \rangle\right].\left[A_j(l) - \langle A_j \rangle\right]}{\sqrt{\sum_{l=1}^{N}\left[A_i(l) - \langle A_i \rangle\right]^2}.\sqrt{\sum_{l=1}^{N}\left[A_j(l) - \langle A_j \rangle\right]^2}},$$

(1)

where $A_i(l)$ is $i^{\text{th}}$ profile with $1 \le l \le N$. Obviously, we could use real-time deformation of the contacts area (deformation in $z$ direction) or the generally shear displacement components- $\Delta \mathrm{u} = (\Delta x, 0, \Delta z)$ where we assumed the interface does not have any shear component in $y$-direction. Additionally, based on a relatively high-curved displacement field, we can build ridge networks. We may see each profile as a separated cycle of a spatial series (i.e., collection of cycles in $x$-direction). To map the obtained series, we define each patch as node. To make an edge between two nodes, relative-high correlated profiles are connected $(C_{ij} \ge r_c)$ with non-direct links. To choose the optimum value of $r_c$ , we notice that the aim is to reach or keep the most stable structures in the total topology of the constructed networks.

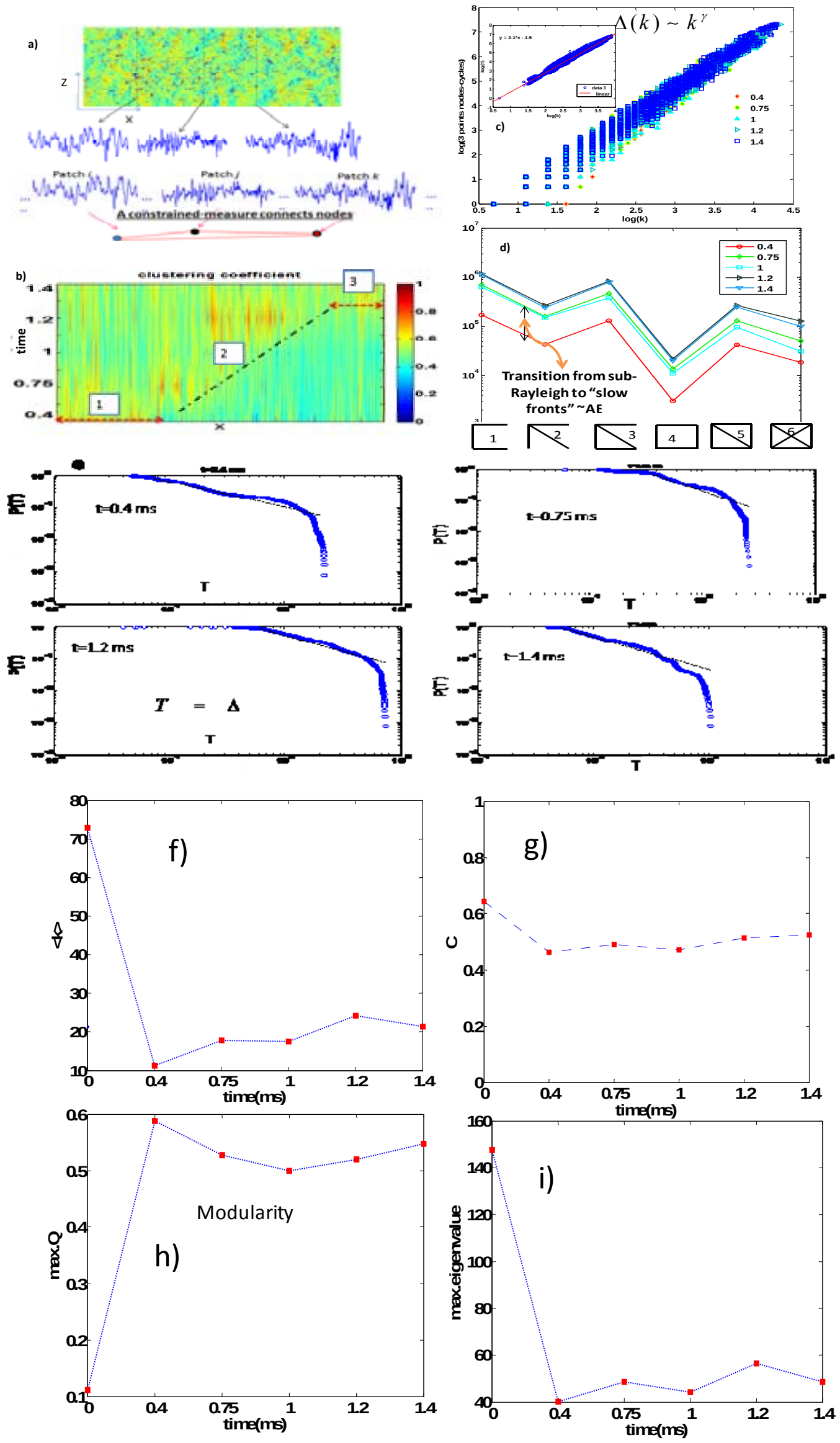

Fig 2. (a) A typical example of transferring contact areas into networks, (b) mapping the dynamic relative contact areas in 2D [1] to networks space and plotting the clustering coefficient as a fraction of triangles reveals the relatively precise rupture speed:three distinctive rupture speeds are compatible with the mean contact area; 1 , 3 corresponds with sub-Rayleigh rupture and 2 is slow rupture; (c) scaling of triangles in obtained networks with number of similar profiles (node's degree) ,expressed with power law relation with universal exponent (~2); (d) distribution of motifs of networks over different time steps shows a non-uniform /universal distribution ;transition from rupture (1) to rupture (2) occurs suddenly which is related to regular acoustic emission (high frequency waveform); (e) distribution of triangles shows power-law distribution –This proves loops congests around "hubs" and This property drives frictional interfaces; (f) node's degree evolution; (g) clustering coefficient versus time; (h) modularity index versus time; (i) maximum eigenvalue during the evolution of the friction network .

Different approaches have been used, such as the density of links, the dominant correlation among nodes, motif density or correlation dimension [18-20]. To choose $r_c$ , we use a nearly stable region in rate of betweenness centrality (B.C) - $r_c$ space, which is in analogy with the minimum value in the rate of the edges' density [20]. The latter method has been used successfully in analyzing time-series patterns in network space [18-20]. We notice finding a nearly stable region in $B.C$- $r_c$ space satisfies the dominant structures of contact patterns. Despite the aforementioned case, most of the recorded photos had insignificant dimension in $z$ direction (<0.2 mm-6mm). Then, naturally with respect to the apparatus, the observed patterns were the longitudinal contact zones. In other words, 2D imaging drastically slows down the rate of the contact area's registration. The measurements were generally 1D, as the integration over the $z$ direction is performed by optical means [2, 4]. To build networks over 1D spatial or temporal contact areas, we use two different methods. The first method is comparing the "closeness" of contact areas; *i.e., if* $|A(x_i,t)-A(x_j,t)|<\xi \rightarrow a_{ij}=1$ where $a_{ij}$ is the component of the connectivity matrix. We use a similar procedure ($B.C$-$\xi$ space) to select the threshold level. The second method is based on transferring time-series to network space by using a standard delay-coordinate embedding method [20, 32-34]. Let us assume a spatial-contact areas series $A(ix)(i=1,2,...,N)$ where $x$ is the size of minimum pixel (in shear direction) and $N$ is the number of pixels (sample size). A sequence of phase-space vectors is constructed using a standard delay-coordinate embedding method [20,34]:

$$\vec{X}=\{A(kx),A(kx+\lambda),...,A(kx+[m-1]\lambda)\}, \qquad (2)$$

where $\lambda$ is the delay space (in analogy with delay time), $m$ is the embedding dimension and $k=1,2,...,N$ ($N$ is the number of vectors in the constructed phase space). To choose the value of $\lambda$ , we use the correlation integral method proposed in [35]. To choose $m$, we use the false nearest neighbors' method for the phase space, which successfully presents the unfolding of orbits in the phase space [36]. Now, with considering each vector point as a node and comparing the cross-correlation of vectors, a link is made if

$$C_{ij}=\frac{\sum_{l=1}^{N}[X_i(l)-<X_i>].[X_j(l)-<X_j>]}{\sqrt{\sum_{l=1}^{N}[X_i(l)-<X_i>]^2}.\sqrt{\sum_{l=1}^{N}[X_j(l)-<X_j>]^2}}>\zeta$$

(the same procedure or edges density was used to choose the threshold level).

To proceed, we use several characteristics of networks. Each node is characterized by its degree $k_i$ (number of links connected to that node) and the clustering coefficient. Clustering coefficient (as a fraction of triangles (3 point loops/cycles) is $C_i$ defined as $C_i = \frac{2T_i}{k_i(k_i - 1)}$ where $T_i$ is the number of links among the neighbors of node *i*. Then, a node with *k* links participates on *T(k)* triangles. With respect to clusters, groups or communities in networks, we are interested in detecting modules and their possible role in conducting frictional interfaces. Furthermore, based on the role of a node in the modules of network, each node is assigned to its within-module degree (*Z*) and its participation coefficient (*P*). High values of *Z* indicate how well-connected a node is to other nodes in the same module, and *P* is a measure to well-distribution of links of the node among different modules [22]. To determine the modularity and partition of the nodes into modules, the modularity *M* (i.e., objective function) is defined as [22]:

$$M = \sum_{s=1}^{N_M}[\frac{l_s}{L} - \left(\frac{d_s}{2L}\right)^2], \qquad (3)$$

in which $N_M$ is the number of modules (clusters), $L = \frac{1}{2}\sum_i k_i$, $l_s$ is the number of links in module and $d_s = \sum_i k_i^s$ (the sum of nodes degrees in module *s*). Using an optimization algorithm (here we use the Louvain algorithm [23]), the cluster with maximum modularity is detected. To describe the correlation of a node with the degree of neighboring nodes, an assortative mixing index is used:

$$r_k = \frac{<j_l k_l> - <k_l>^2}{<k_l^2> - <k_l>^2} \qquad (4)$$

where it shows the Pearson correlation coefficient between degrees $(j_l, k_l)$ and <•> denotes average over the number of links in the network.

## 3. RESULTS AND FORMULATION: FRICTION LAW WITH NETWORKS

We start with 2D interfaces, monitored at discrete time steps [0, 0.4, 0.75, 1, 1.2, and 1.4] *ms* (Figure 1a) [1]. Transferring X-patches (then perpendicular to shear direction) to networks and plotting the clustering coefficient revealed three distinct patterns of rupture evolution (Figure2.b), which are comparable with the previous results [1-2]. The three patterns correspond with sub-Rayleigh (1 and 3 in Figure2.b) and slow rupture (2 in Figure2.b). In other words, the movement of the rupture tip is followed by the fast variation of the clustering coefficient. This may be assumed to be a coupling of the dynamic and topology of the system. Furthermore, considering 3 points cycles (*T*-triangle loops) versus node's degree from 0.4-1.4 ms shows a power law scaling (Figure2.c):

$$T(k) \sim k^{\beta}, \qquad (5)$$

where the best fit for the collapsed data set reads $\beta \approx 2.1 \pm 0.4$ (we call the coupling coefficient of local and global structures). A similar scaling law is obtained for other types of sub-graphs (Figure 4). Peculiarly, Euclidean distance (instead of Eq1.) yields the same scaling power law with different coefficient and power (Figure 5). Thus, our analysis of aperture patterns in a discrete slip measurement (over 20mm shear displacements) in rock samples revealed the same scaling law with a very close coupling coefficient [18]. With some mathematical analysis, one can show that adding *m* edges increases the number of loops with $\beta^2 m^{\beta}$, indicating a very congested structure of global and local sub-graphs during shear rupture. Also, we notice $C(k) \sim 2k^{\beta-2}$ so that for $\beta < 2$, a hierarchical structure is predicted [25]. Hereafter, we develop some non-dimensional equations. We assume the variation of the clustering coefficient around the rupture zone is proportional with the variation of shear load, i.e., $\frac{\partial f_s}{\partial t} \sim -(\frac{\partial C(k)}{\partial t})$. Consequentially, we obtain $\frac{\partial f_s}{\partial t} \sim k^{\beta-2}(2-\beta)\frac{\partial \ln k(t)}{\partial t}$ and in terms of loops it reads $\frac{\partial f_s}{\partial t} \sim 2(2-\beta)T(k)^{(1-\frac{3}{\beta})}$. With $\varepsilon \equiv k^{\beta-2}(2-\beta)$ we obtain $f_s \sim \varepsilon \ln k(t)$. The latter relation is similar to rate and state friction law ($f_s \sim \ln\theta$ where $\theta$ is the state parameter). Consequently, one may present a friction-network law. We also estimate the speed of the rupture's fronts around the rupture zone with (Figure 2b): $\frac{\partial^2 C(x,t)}{\partial x \partial t} \sim -1/v_{front}$ which states that the temporal-spatial gradient of the triangles' fraction correlates with the inverse of the rupture fronts' speed. Next, we look at the distribution of loops (Figure2.e). Irrespective of the different type of rupture modes in the considered time interval, a power law satisfies the obtained distributions:

$$P(T(k)) \sim T^{-\gamma}, \qquad (6)$$

in which $\gamma \approx 2.0145$. The power law nature of loops shows the aggregation of cycles around "hubs". In other words, during rupture evolution, loops (and generally sub-graphs with loops) are not distributed uniformly. In Figure2.f, we show that friction-networks are assortative, i.e., nodes tend to connect to vertices with a similar degree. Let us transfer (6) into a "hubness" model (as well as Barabasi-Albert model [26]) where we assume hub nodes tend to absorb more loops rather than poor nodes. In this model, the probability of finding a component of the network around hubs is higher than other nodes. Then, it leads to:

$$\frac{\partial T_i(k)}{\partial t} \sim m\frac{T_i}{\sum T_j}, \qquad (7)$$

in which $m$ is a coefficient of growth (or decay). Plugging (5) in (7) yields:

$$\frac{\partial k_i}{\partial t} \sim \frac{m}{\beta}\frac{k_i}{\sum k_j^{\beta}}. \qquad (8)$$

For $\beta = 1$ the model yields scale-free networks. To complete our analysis, we write the state parameter of state and rate friction law [29], as a linear combination of local and global structures. We consider a simplified form of the standard equation for the friction with assumption of nearly constant sliding velocity: $\mu_s \sim \ln\frac{\theta}{D_c}$ in which $\theta$ is the variable describing the interface state and $D_c$ is the characteristic length for the evolution of $\theta$. The state variable carries information about the whole population of asperities [30]. Then it is reasonable to estimate the evolution of the state parameter with the collective behavior of the system's elements. Commonly used empirical laws for evolution of state variable are Ruina's laws for ageing and slipping states [15]. For slip law, it reads: $\frac{\partial \theta}{\partial t} \sim -\frac{\theta}{D_c}\ln\frac{\theta}{D_c}$ . Let us transfer the state variable in the network space as follows (we have considered the first order of the multi-parameters of the Taylor's expansion):

$$\frac{\partial \theta_i(t)}{\partial t} = a\frac{\partial k_i}{\partial t} + b\frac{\partial T_i}{\partial t}, \qquad (9)$$

which indicates that the evolution of the state variable is a superposition of the local and global properties of friction networks. We eventually obtain:

$$\frac{\partial \theta_i(t)}{\partial t} = \frac{\partial k_i}{\partial t}(a + b\beta k_i^{\beta-1}). \qquad (10)$$

With assuming $\beta = 1, \frac{\partial k_i}{\partial t} > 0$ and $a < 0$ , eq.(10) indicates a decaying model for the state parameter in terms of attacking to hubs. Plugging (8) into (10) and assuming $\beta = 2$ leads to:

$$\frac{\partial \theta_i(t)}{\partial t} = \frac{2bmk_i^2}{\sum k_j^2} + \frac{amk_i}{\sum k_j^2}, \qquad (11)$$

which shows the complex non-linear nature of the state parameter. Further developments of (11) and (12) can be done by using non-linear growth models as well as rate equations [27]. It is noteworthy that the first part at the right side of Eq. 11 shows friction networks follow a gel-like state, in which a condensation of nodes occurs. The condensation means that the whole interface shows similar characteristics. In this case (or when $\frac{a}{b} \to 0$ ), a single node is connected to almost all the nodes in the friction network. This behavior is biased with the second term if the weight of global evolution is significant. Let us consider a case in which a possible condensation of loops occurs; i.e., $\frac{\partial T_i(k)}{\partial t} \sim m\frac{T^{\nu}_i}{\sum T_j^{\nu}}$ where we assume $\nu = 2, \beta = 1$. Substituting in (10), we obtain: $\frac{\partial \theta_i(t)}{\partial t} = \frac{m(a+b)k_i^2}{\sum k_j^2}$ . To solve the later relation, we estimate $\frac{\partial k_i(t)}{\partial t} = \frac{m(a+b)k_i^2}{\sum k_j^2} \approx \frac{m(a+b)k_i^2}{t^2}$ which eventually results $k_i(t) \sim \frac{t}{m(a+b)}$. We notice this is a super-linear behavior in terms of the evolution of the possible friction network.

Let us conclude the possible combinations of power-law scaling and edges or loops growth. Instead assuming (7), take $\frac{\partial k_i}{\partial t} \sim \frac{k_i^{\delta}}{\sum k_j^{\delta}}$ and $T(k) \sim k^{\beta}$, eventually it leads to the following results:

$$T_i(t) \sim \begin{cases} t & (\delta,\beta) = (2,1) \\ \frac{1}{3}t^{3/2} & (\delta,\beta) = (2,1/2), \\ (\ln t)^{-2} & (\delta,\beta) = (2,2) \end{cases} \qquad (12)$$

and for $\frac{\partial T_i}{\partial t} \sim \frac{T_i^{\nu}}{\sum T_j^{\nu}}$ It reads to:

$$T_i(t) \sim \begin{cases} t & (\nu,\beta) = (1,1) \\ e^{\frac{t^3}{6}} & (\nu,\beta) = (1,\frac{1}{2}). \\ e^{4\sqrt{t}} & (\nu,\beta) = (1,2) \end{cases} \qquad (13)$$

From Eq. (12), we conclude that two components of the friction network cannot show gel-like behavior at the same time. In other words, if the friction network's links are connected to a single hub, then loops cannot be condensated. In the case of aggregation of loops around hubs (Eq.13), a super-aggregation of loops grows when $0 < \beta < 1$ . We conjecture generally decreasing $\beta$ corresponds with the most unstable status of the rupture mode. A different but much more physical approach is the idea of the growth of friction networks with respect to the fitness of each node (~profile), proportional to rate of dissipated energy of each patch. Smaller fitness is assigned to the patches with high energy (highly unstable) nodes, such that [28, 37] $\Gamma_i \sim -v_i\eta_i^{\theta}$, in which $\eta_i$ is the fitness parameter and $v$ is related to the velocity (slip rate) of the interface and generally can be assumed as an external energy induced to the friction network.
Notably, the presented fitness relation is compatible with fracture energy (dissipation rate of energy), which states $\Gamma \sim (1 - v/c_R)$. Let us assume the evolution of edges is given by (this can be assumed as a constitutive equation): $\frac{\partial k_i}{\partial t} = \frac{\eta_i k_i}{\sum_l \eta_l k_l}$. Then, considering Eqs. (5-6) and assuming that the distribution of $\Gamma_i$ is proportional to the density of loops, for a fixed velocity for all nodes, one can achieve: $\frac{\partial k_i}{\partial t} \sim \frac{k_i^{\kappa\theta+1}}{\sum_l k_l^{\kappa\theta+1}}$ where $\Gamma \sim T^{\vartheta} \sim k^{\kappa}; \kappa = \vartheta\beta$ ($\vartheta$ is a constant parameter). The obtained simple relation is a non-linear kernel form of the network's evolution and has been addressed in [26-27], where, for different situations of $\kappa\theta + 1$, different regimes of the evolution are predicted. For instance, if $\kappa\theta \geq 0$, then the network shows a super-linear facet. As a conclusion, we tried to express the state parameter of the friction law in terms of the node 's degree, where we considered the role of dissipation energy rate and loops in our extensions.

Considering the spatial gradient of the state variable's rate and recalling the relation of rupture speed and clustering coefficient requires (writing Eq. 8 in terms of $k$ and $C$ and plugging $\frac{\partial^2 C(x,t)}{\partial x \partial t} \sim -1/v_{front}$): $1/v_{front} \sim -\chi(k) \frac{\partial^2 \theta(x,t)}{\partial x \partial t}$ where $\chi(k)$ is a function of node's degree and coupling coefficient. To confirm this relation using the same methodology from [11-12], we estimate front's speed. Assuming $\frac{\partial \bullet}{\partial t} = -v_{front} \frac{\partial \bullet}{\partial x}$ and recalling Ruina's law with the assumption of unity rate of displacement $\frac{\partial \theta}{\partial t} = 1 - \frac{\theta}{D_c}$, then we obtain $v_{front} \frac{\partial^2 \theta(x,t)}{\partial x \partial t} + \frac{\partial v_{front}}{\partial t} = \frac{-1}{D_c}$ which is comparable with our prediction of the front's speed. Next, we investigate the frequency of the sub-graphs. Sub-graphs are the nodes within the network with special shape(s) of connectivity together. The relative abundance of sub-graphs has been shown to be an index of a networks' information processing functionality. A motif is a sub-graph that appears more than a certain amount. It can be defined more precisely using Z-scores, which compare motif distribution with corresponding random networks [38-39]. A motif of size $k$ (containing $k$ nodes) is called a $k$-motif (or generally sub-graph). We plot sub-graphs (*i.e.,* motifs) distributions (Figure 2d) that indicate a super-family phenomenon. A similar trend in all the rupture speeds indicates the universality of energy flow in frictional interfaces that are characterized by friction laws. We examined this universality by measuring the contacts in terms of discrete displacement intervals and confirmed the obtained results [19]. Remarkably, transition from sub-Rayleigh to slow rupture is correlated with a distinct spike in all types of the 4-points sub-graphs. Considering the variations of edges density $<k>$ at the monitored interval (where $10 \leq k \leq 30$) and the abrupt change in loops, which is in order of one magnitude, we conclude that the fluctuation of the coupling coefficient induces remarkable growth of sub-growth. One can confirm increasing sub-graphs is consistent with our formulation and observation (for 1D-friction patterns).

## 4. 1D INTERFACES: MODULARITY, ASSORTATIVITY AND SELF-ORGANIZED CRITICALITY

Following 1-D patterns of contact areas, we map 1D-net contact areas into friction networks using a closeness metric (Figure 3). According to [3] – and Figure 3b-, the contact areas feature three remarkable evolutionary stages, as follows: 1) detachment phase, 2) rapid slip, and 3) slow slip. We notice, in this case, that the first phase accompanies around 10% of contact area variation, while phase 2 represents 2-3% of the changes in contacts. Notably, the evolution of pure contact areas –in this space– does not reveal any more information about rupture speed or any other possible mechanism behind the fracture evolution. Employing the first method to construct friction networks unravels some unique characteristics of the rupture fronts. In the following, we present a new perspective to describe the mechanisms of slow or fast ruptures. The corresponding friction networks' parameters, such as assortativity index and maximum modularity (Q) [31], indicate more features of the rupture speed. Each front passage encoded as the remarkable spikes in assortativity index or maximum modularity (Figure 3b, f). We also confirmed the coupling coefficient dramatically drops during the transition of fronts. Based on our formulation, the rapid drop of the coupling coefficient induces the fast variation of normal force (proportional to strength of asperities).
Moreover, the maximum variations of assortativity and modularity at the first phase are ~50% and ~30%, respectively. A significant spike in transition to the phase (1) is followed by a nearly large stationary value of assortativity, where the transition time is comparable to the propagation of sub-Rayleigh's front. Generally, our observations showed that rapid sub-Rayleigh fronts encoded as sudden spikes in the assortativity index, comparable with the radiated energy and acoustic emissions. Remarkably, the slow rupture stage is generally represented by the monotonic growth of modularity, while the variation of assortativity occurs in two distinct categories: stable and non-stable slow rupture. Stability in assortativity occurs in two different stages, one is before sliding and another one takes place in the net-movement of the interface. However, the mechanisms of the similar stable-assortative stages are different. The first stationary interval accompanies the growth of modularity, while the continuous decaying of links happens. The second interval, which occurs with the same growth of assortativity (>20%), shows a decreasing trend in modularity. We conclude after a rapid-strong sub-Rayleigh front, a silent period is encoded into the system, followed by a fast transition to the next stage. Indeed, transition to phase 3 occurs with ~25% fast-increment of assortativity index, while the variation of the contact areas is only less than 4%. In other words, during this small time-interval (~60-100μs) what controls the behavior of the interface is the pattern (shape) of contacts—rather than pure contact areas where we guess the central rich-hubs drive the system. Immediately, it follows a relatively long stationary assortativity, indicating a stable mechanical characteristic of the interface (possible stable-low sliding rate). With respect to our observations, we hypothesize that stable and unstable-slow ruptures carry high and small localization, respectively, while sub-Rayleigh ruptures hold high localization in terms of the evolution's rate of the assortativity. Obviously, irrespective of slow rupture modes, the growth of modularity is the unique characteristic of slow rupture . Phase 3, clearly is characterized by a long term of near stability in assortativity, net-contact areas, and node's degree (with a small ~3% growth in modularity, clustering coefficient are observed). We have confirmed these results using two more data sets. It is worth mentioning that the maximum eigenvalues of the Laplacian matrix show approximately the same trend of degree correlation (Figure.3g).

Remarkably, stable slow rupture is associated with a low level of synchronizability. The transition to an unstable period is coupled with a soft increase in synchronizability, followed by a successive decline. Interestingly, a large spike in synchronization occurs during the interface's transition to the sliding stage, inducing ~10% decrease in synchronizability and then inducing significant instability. We conclude that the transition from stick to slip requires relative minimum synchronization in the corresponding friction networks, showing a fast release of energy and rapid slip rate. One may interpret the localization of energy in terms of friction networks' synchronizability, stating that high synchronizability lowers the localization. As one can follow, we indirectly prove the self-organized criticality (SOC) in frictional interfaces with regard to networks parameters. Since the 1980s, the theory of SOC over the time-series of earthquakes (either in laboratory scale or large scale) has been suggested to play a significant role in diverse avalanche-like behavior of materials (or "crackling noise") [40]. In our case, the interface is evolved in a way that the minimum variation of contact areas reaches the optimum quasi-stable regime (the best possible robust contact patterns).

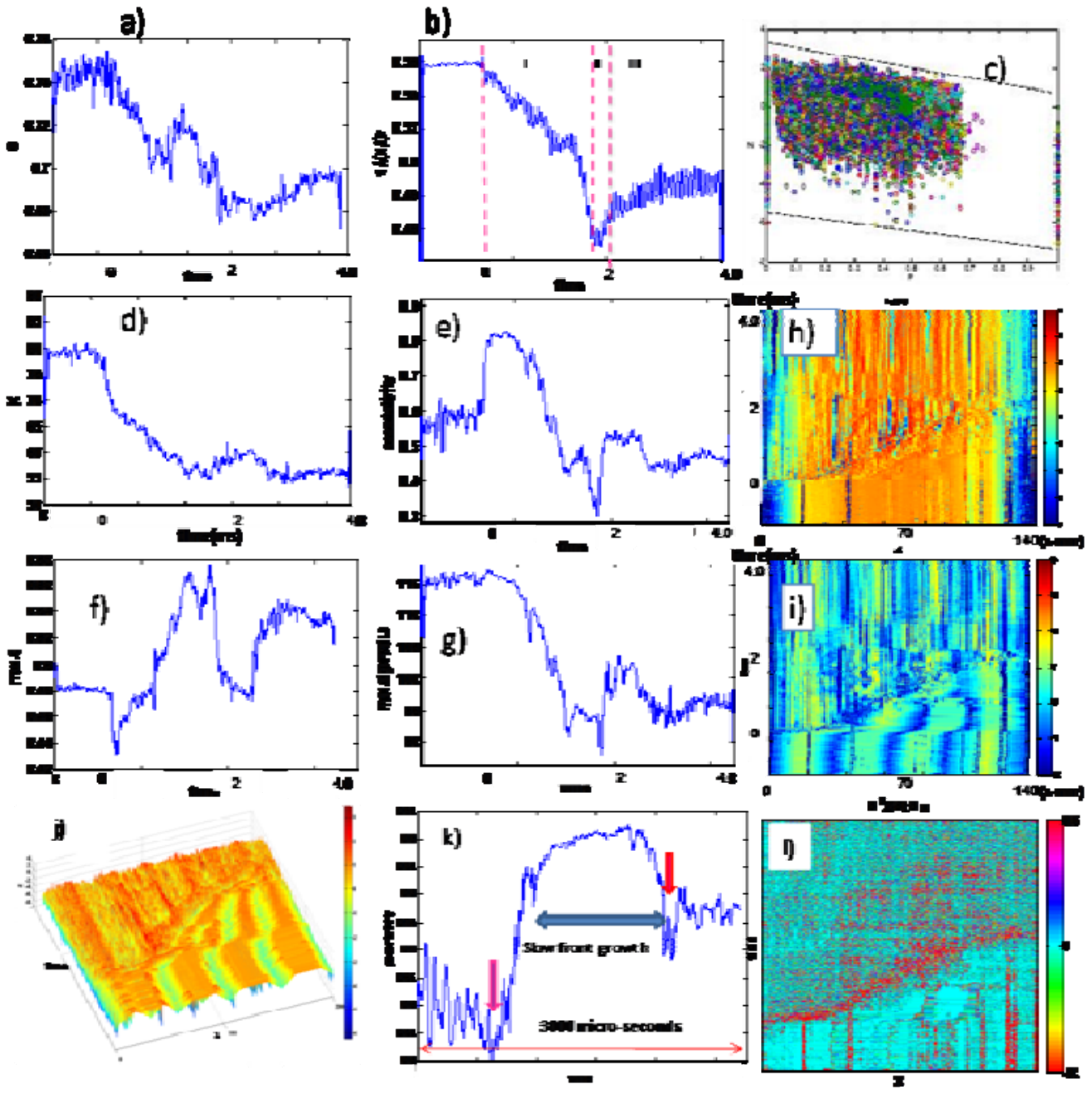

Fig 3. Transferring 1D contact areas [1] to networks with a spatially constrained metric; (a) the clustering coefficient versus time (~4ms), (b) the real-time changes of contact areas (c) each element of the interface-network is mapped into modularity space (participation factor of each edge and within module degree); (d) the node's degree versus time (~4ms); (e) the assortativity coefficient versus time (~4ms) shows all of networks are assortative; (f) maximized Q as an index to modularity versus time; (g) the maximum eigenvalues of Laplacian of the node's degree against time indicate possible synchronization frictional interfaces' elements; (h) Betweenness Centrality (B.C.) –as a measure of a node's centrality in a network-indicates the clear rupture transitions and possible periodic nature of the interface (characteristic length in friction law); Here we have plotted the natural logarithmic variation of B.C. and (i,j) the variation of within module degree (Z) precisely shows intersonic fronts and the details of fronts; the periodic nature of modules clearly affected by detachment fronts. In panel (j), we have shown the results for another case of mixed rupture (contact areas data set courtesy of S.M. Rubinstein and O. Ben-David); (k) the assortativity variation corresponding to panel (j); the arrows show the onset of fast-strong fronts; (l) the changes in x-t- $\frac{\partial^2 Z}{\partial t \partial x}$ show an obvious index to rupture path and rupture's speed.

Interestingly, modeling networks with variable assortativity have been shown to be affected by the correlation of nodes, such that scale-free networks with larger assortativity coefficients are more robust against "attack" [41-42]. Specifically, assigning the period of rapid slip to the cooling down (i.e., thermal) process will not explain the spikes in the correlation of node's degree [3]. Mapping the nodes of friction networks in modularity space, defined by the within-module index (Z) and participation coefficient (P), revealed collapsing of all nodes in a certain range of P and Z (Figure 3C). We confirmed the universality of observed patterns in P-Z space over different case studies, either real time contacts (1 or 2 D) or aperture friction networks.

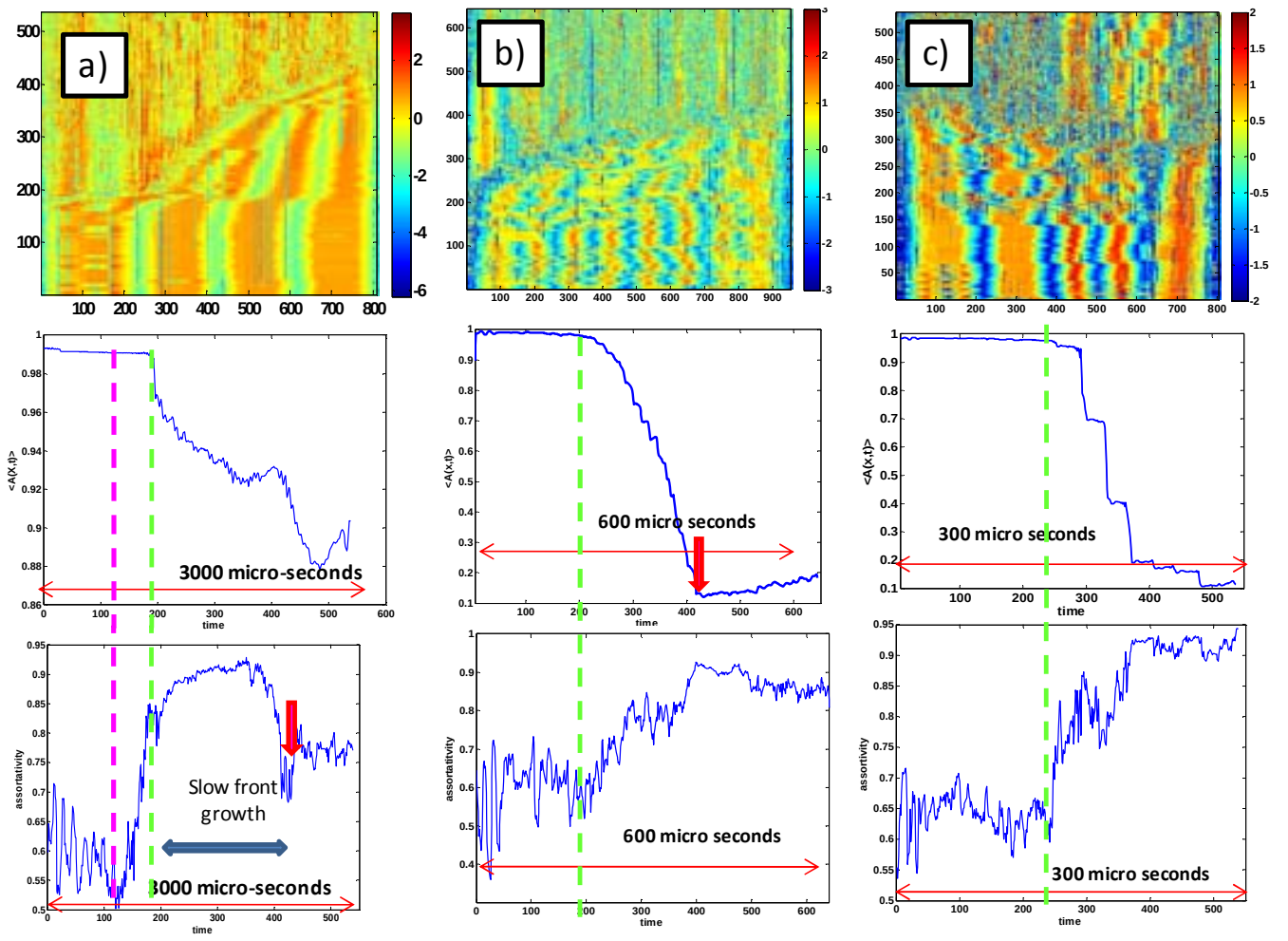

Fig 4. a) Mixture ruptures: within module degree evolution (the vertical axis shows time [~3000micro-secons] and the horizontal axis represents the X-axis), the mean relative-contact area and assortativity index (left hand panel); b) within module degree evolution of pure sub-Rayleigh rupture; c) within module degree evolution of pure super-shear rupture (contact areas data set courtesy of O. Ben-David [46]). The broken-green lines show the onset of instability in the interface.

Following [21], we divide the P-Z space into 7 sub-regions: $R_1,...,R_7$ which are based on the role of non-hubs and hubs nodes. The general categorization of the P-Z parameter space is as following:

$$Non-hubs:\{R_1(P\le 0.05),R_2(0.05<P\le 0.62),R_3(0.62<P\le 0.80),R_4(P>0.80)\}\cap\{Z<2.5\},$$
$$Hubs:\{R_5(P\le 0.3),R_6(0.3<P\le 0.75),R_7(P>0.75)\}\cap\{Z\ge 2.5\}.$$

where each region nominates a unique characteristic of rich or poor nodes with respect to other modules . Our results for the 8 case studies show the probability of finding a profile in $R_4$ and $R_7$ is very unlikely and the most congestion of profiles happens in $R_1$, $R_2$ and $R_3$. A few nodes occupy $R_5$, indicating the role of hub nodes with the majority of connections within their module. Amazingly, plotting the evolution of the within module index in *x-t*-Z*(x,t)* shows a periodic nature (Figure 2*i*). We obtained the periodic property of Z*(x,t)* in other real-time contact areas and aperture-friction networks. We believe the periodicity of the within-module degree and the participation coefficient —with respect to the nearly invariant nature of spatial-periodicity— are related to the characteristic length ($D_c$) of the interface. Interestingly, in this parameter space, clear super-shear fronts can be observed. Further studies will highlight the network-signatures of super-shear fronts (rupture). As in the previous case, we can develop the evolution of the state parameters with respect to the modularity parameters: $\frac{\partial\theta}{\partial t}=c_z\frac{\partial Z}{\partial t}+c_P\frac{\partial P}{\partial t}$. We have (see figure 3.l) $\frac{\partial^2 Z}{\partial x\partial t}\sim 1/v_{front}$ then $\frac{\partial\theta}{\partial t}=\frac{\partial Z}{\partial t}(c_z+c_P\frac{\partial P}{\partial Z})$ and we find: $\frac{\partial^2\theta}{\partial x\partial t}\sim v_{front}^{-1}(c_z+c_P\frac{\partial P}{\partial Z})+c_P\frac{\partial Z}{\partial t}\frac{\partial^2 P}{\partial x\partial Z}$, in which $c_z$ and $c_P$ are appropriate constant variables (as well as the weights of modularity parameters) . We estimate the slope of the upper (or lower) bond of the P-Z space with a constant gradient: $\frac{\partial P}{\partial Z}\approx -\delta$ . Hence, it leads to: $(\frac{1}{c_z-\delta c_P})\frac{\partial^2\theta}{\partial x\partial t}\sim 1/v_{front}$, denoting that increasing $\delta$ (dropping the slope of P-Z ) induces decreasing of $v_{front}$.

In figure 4, we have presented the results of friction networks over recent real-time contact measurements [46] for three different types of the ruptures, namely: 1) mixed rupture (sub-Rayleigh +slow), 2) sub-Rayleigh, and 3) super-shear. As one can follow, the within-module degree follows spatial-periodic patterns which are distinguishable before the arrival of rupture fronts and without respect to the type of the rupture. We confirmed our previous approach with respect to the signatures of the node's correlation during the propagation of the rupture.

To complete our analysis of 1-D patterns, we transfer spatial patterns in a certain time to network space, using a standard delay-coordinate embedding method. This method possibly will be useful in analyzing waveforms where the limited waveforms are recorded through seismograms or transducers (see figure 7 and [19]). Also, it should be noted that we can change from a spatial-series to a temporal form, i.e. time-series. In Figure 5, the results of this method over another 1D frictional interface have been shown. Here, the initial transition from sub-Rayleigh to slow rupture occurs very fast, while the second transition leading to net-movement of the interface takes a longer time interval. Interestingly, we again observe the two aforementioned scaling laws (Figure 5b, 5c). The assortativity also follows nearly the same patterns as the closeness metric.

Remarkably, before the arrival of an extremely fast front –directing to sliding- a distinguishable dropping of assortativity is encoded (i.e., unstable slow rupture). It should be emphasized that, in this case and with this metric, we do not see the same characteristic of modularity (*Q*) that we observed in closeness. Analysis of the data set in the temporal form and with the standard delay-coordinate embedding method unraveled interesting behavior in terms of the different networks' parameter space (Figure 6).

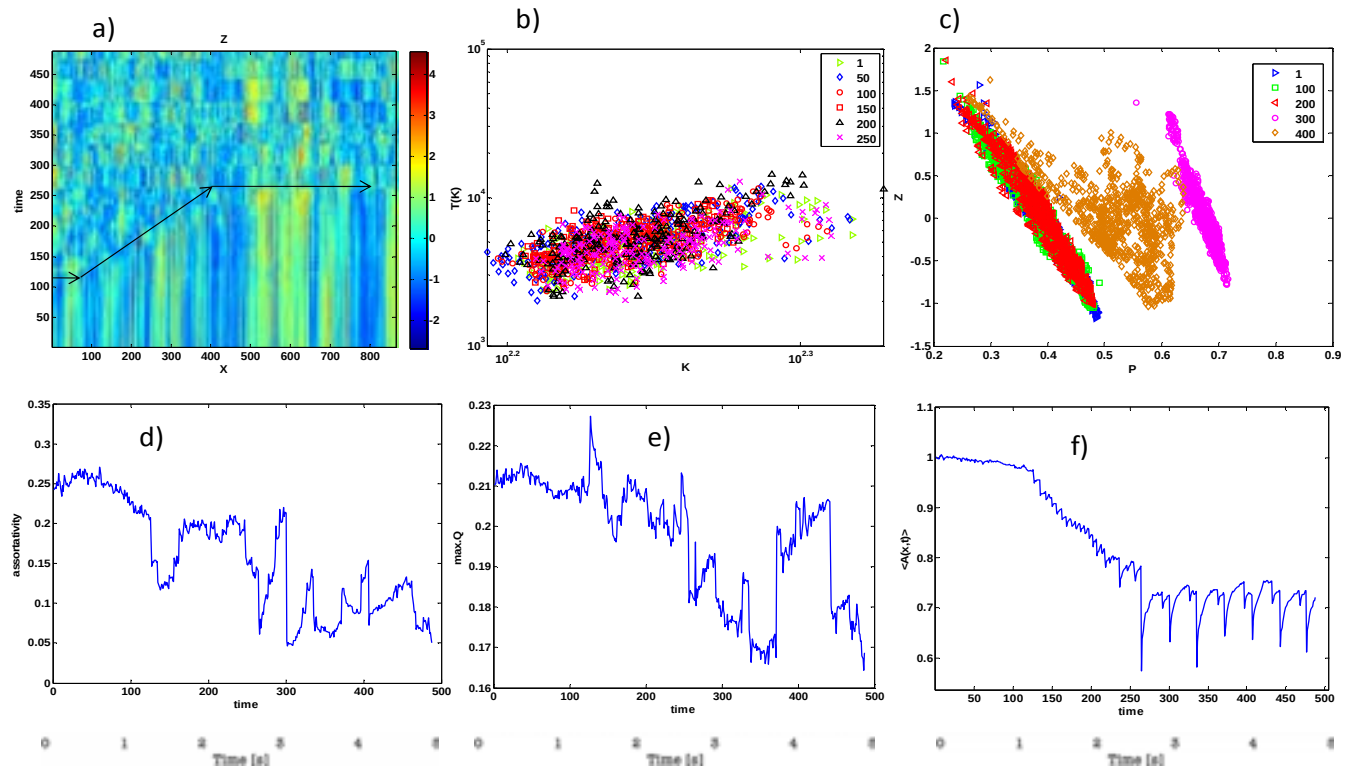

Fig **5**. Results of the time delay method for each time step (i.e., spatial –series) and converting to network space based on the correlation of the embedded series (contact areas data set courtesy of S. Maegawa); a) Z(x,t)-x-t space parameter; b) scaling loops with node's degree; c) P-Z space and $\frac{\partial P}{\partial Z}\approx -\delta$ d) assortativity vs. time; e) maximum modularity versus time and f) variation of contact areas in the sheared dry interface.

For instance, averaging over spatial nodes for *B.C* and *k* (Figure 6b) show a nearly oscillator like behavior which limiting to an attractor. Interestingly turning points in this space or *c-k* parameter space correspond with transition to another rupture's mode (i.e., front's speed). For example the slow rupture in *c-k* parameter space is given by: $<C>_x \simeq 10^{-3}<k>_x +0.62$ or with an arch of a circle for *B.C* –*K* space: $(\ln<B.C>_x -6.5)^2+(<k>_x -100)^2 \approx 1600$. The aforementioned method applied to the time series is analogous to the interpretation of acoustic waveforms, where hundreds of transducers were used to record the evolution of the interfaces. We have used the latter method to analyze the results of continuously recording a smooth fault with 13-16 transducers while the fault surface was completely 2D (more results will be reported elsewhere but see [43]).

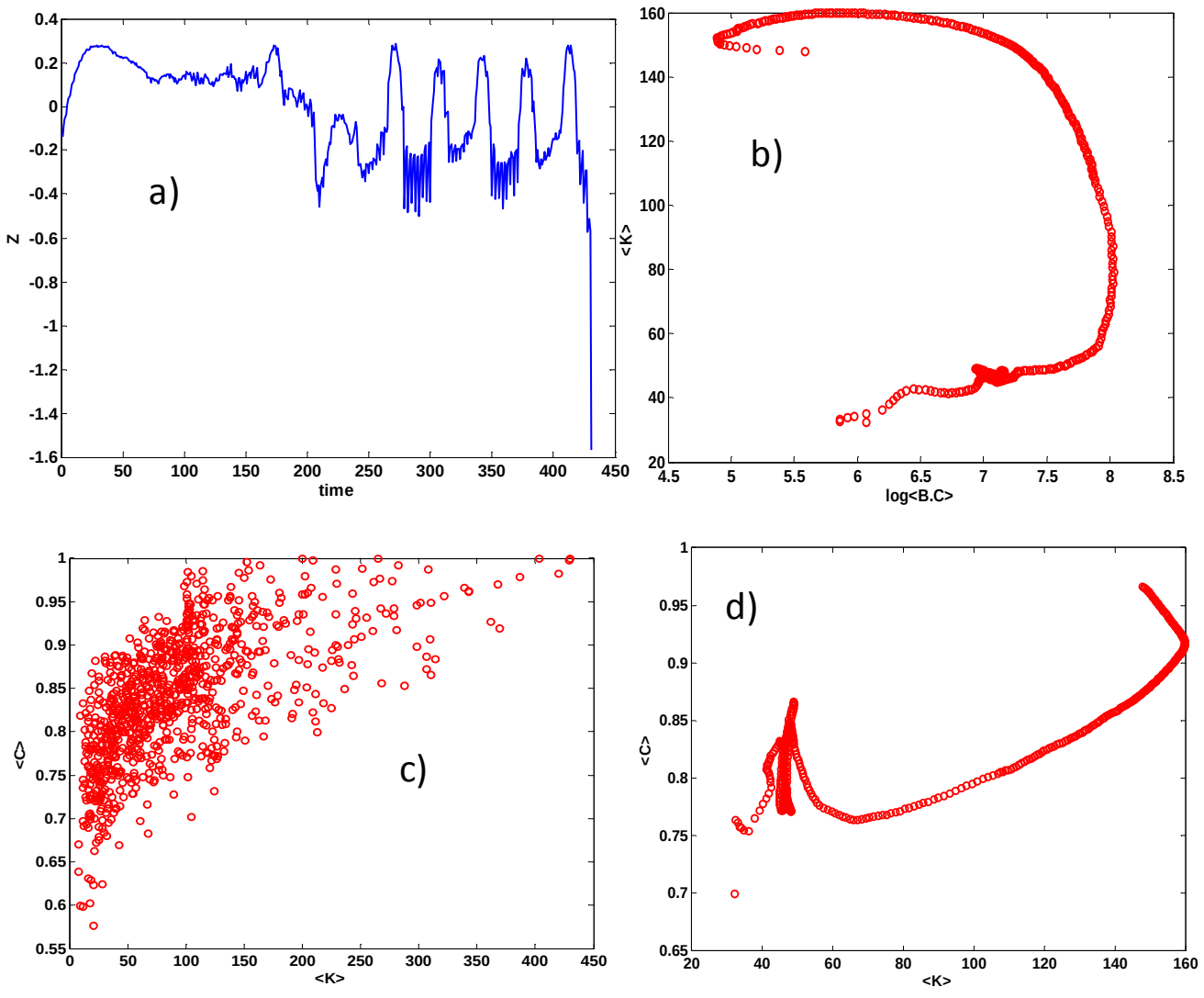

Fig 6. Results of the time delay method for each time step (i.e., temporal –series) and converting to network space based on the correlation of embedded series (contact areas Data set courtesy of S. Maegawa); a) averaging Z(x,t) over space parameter; b) Phase-space like behavior of log<B.C>-<k>; c) averaging over time in the C-k parameter space and d) averaging over space in the C-k parameter space.

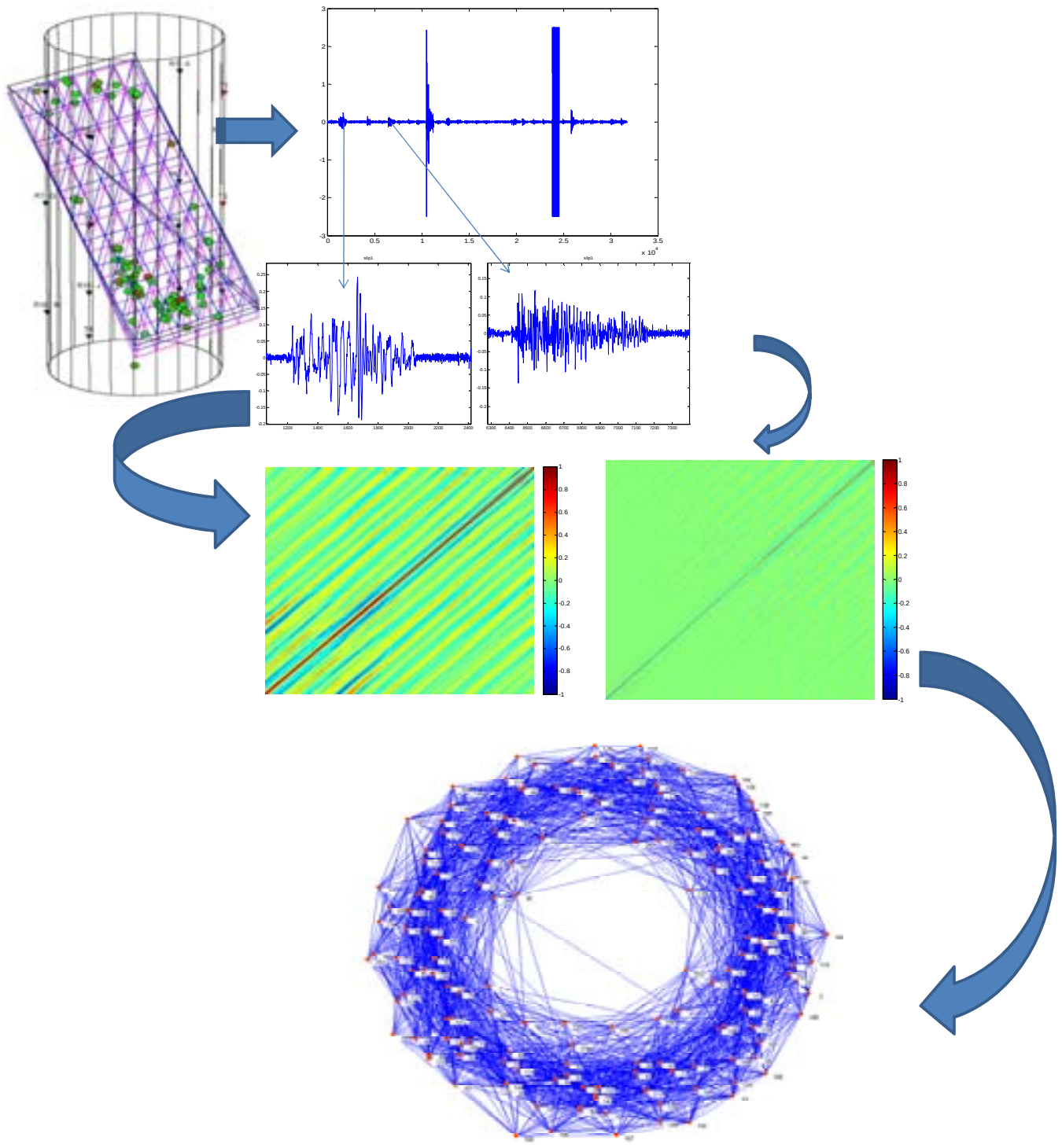

Fig.7. The methodology used to extract friction networks from the acoustic signals of the evolution of a smooth fault (interface). The figure shows two different acoustic waves corresponding to regular shear and slow rupture. To analyze waveforms and to construct networks, the standard-delay time method was used in conjunction with the thresholded adjacency matrix.

## 5. CONCLUSIONS

As a conclusion to our study, we introduced friction networks over dynamics of different real time contact areas. Based on our solid observations we formulated a probabilistic frame to evolution of the state variable in terms of friction networks. Moreover, we confirmed slow ruptures, generally, hold small localization while regular ruptures carry high level of energy localization. Significantly, sub-graph distribution was scaled with possible transition of rupture's speed. We also introduced two new universalities with respect to evolution of dry frictional interfaces: scaling of local and global characteristics and occupation of certain regions of modularity parameter space. Our results showed how relatively high correlated "elements" of an interface can reveal more features of the underlined dynamics. We proposed assortativity as an index to correlation of node's degree can completely uncover acoustic features of the interfaces.

Our formulation can be coupled with elasto-dynamic equations to give a complete picture of frictional interfaces behavior. Furthermore, one can also model and analysis the friction networks in the frame of complex network theory.

***

We would like to acknowledge and thank J. Fineberg and O. Ben-David (The Racah Institute of Physics, Hebrew University of Jerusalem, Israel) ,S. Maegawa (Graduate School of Environment and Information Sciences, Yokohama National University, Japan) and B.D.Thompson (University of Toronto) who provided the data set employed in this study.